\documentclass[12pt]{article}
\usepackage{array}
\usepackage{graphicx}
\usepackage{amssymb}
\usepackage{amsmath}
\input{epsf}
\usepackage{cite}
\def\@fmsl@sh#1#2#3{\m@th\ooalign{$\hfil#1\mkern#2/\hfil$\crcr$#1#3$}}
 \def\eq#1\en{\begin{equation}#1\end{equation}}
\def\s[#1,#2]{[#1\stackrel{\star}{,}#2]}
\def\sx[#1,#2]{[#1\stackrel{\star_{x}}{,}#2]}

\newcommand{\nc}{\newcommand}
\nc{\beq}{\begin{equation}}
\nc{\eeq}{\end{equation}}
\nc{\beqa}{\begin{eqnarray}}
\nc{\eeqa}{\end{eqnarray}}

\def\bc{\begin{center}}
\def\ec{\end{center}}

\def\to{\rightarrow}

\def\gsim{\mathrel{\mathpalette\atversim>}}

\def\bc{\begin{center}}
\def\ec{\end{center}}

\def\gsim{\mathrel{\rlap{\lower4pt\hbox{\hskip1pt$\sim$}}

    \raise1pt\hbox{$>$}}}       %greater than or approx. symbol

\def\gsim{\mathrel{\rlap{\lower4pt\hbox{\hskip1pt$\sim$}}
    \raise1pt\hbox{$>$}}}       %greater than or approx. symbol

\begin{document}
\makeatletter
\def\fmslash{\@ifnextchar[{\fmsl@sh}{\fmsl@sh[0mu]}}
\def\fmsl@sh[#1]#2{%
  \mathchoice
    {\@fmsl@sh\displaystyle{#1}{#2}}%
    {\@fmsl@sh\textstyle{#1}{#2}}%
    {\@fmsl@sh\scriptstyle{#1}{#2}}%
    {\@fmsl@sh\scriptscriptstyle{#1}{#2}}}
\def\@fmsl@sh#1#2#3{\m@th\ooalign{$\hfil#1\mkern#2/\hfil$\crcr$#1#3$}}
\makeatother
%\baselineskip 24pt

%%%%%%%%%%%%%%%%%%%%%%%%%%%%%%%%%%%%%%%%%%%%%%%%%%%%%%%%%%%%%%%%%
%%%
%%%                      TITLE PAGE
%%%
%%%%%%%%%%%%%%%%%%%%%%%%%%%%%%%%%%%%%%%%%%%%%%%%%%%%%%%%%%%%%%%%%
\thispagestyle{empty}
\begin{titlepage}
\boldmath
\begin{center}
  \Large {\bf Unitarity bounds on  low scale quantum gravity}
    \end{center}
\unboldmath
\vspace{0.2cm}
\begin{center}
{
{\large Michael Atkins}\footnote{m.atkins@sussex.ac.uk} and 
{\large Xavier Calmet}\footnote{x.calmet@sussex.ac.uk}
}
 \end{center}
\begin{center}
{\sl Physics and Astronomy, 
University of Sussex,  \\ Falmer, Brighton, BN1 9QH, UK 
}
\end{center}
\vspace{\fill}
\begin{abstract}
\noindent
We study the unitarity  of models with low scale quantum gravity both in four dimensions and in models with a large extra-dimensional volume. We find that models with low scale quantum gravity have problems with unitarity below the scale at which gravity becomes strong. An important consequence of our work is that their first signal at the Large Hadron Collider would not be of a gravitational nature such as graviton emission or small black holes, but rather linked to the mechanism which fixes the unitarity problem. We also study models with scalar fields with non minimal couplings to the Ricci scalar. We consider the strength of gravity in these models and study the consequences for inflation models with non-minimally  coupled scalar fields. We show that a single scalar field with a large non-minimal coupling can lower the Planck mass in the TeV region. In that model, it is possible to lower the scale at which gravity becomes strong down to 14 TeV without violating unitarity below that scale.
\end{abstract}  

\end{titlepage}

%\pacs{}

%%%%%%%%%%%%%%%%%%%%%%%%%%%%%%%%%%%%%%%%%%%%%%%%%%%%%%%%%%%%%%%%
%%%
%%%                     INTRODUCTION
%%%
%%%%%%%%%%%%%%%%%%%%%%%%%%%%%%%%%%%%%%%%%%%%%%%%%%%%%%%%%%%%%%%%

\newpage

\section{Introduction}
The energy scale at which quantum gravitational effects become strong could be much lower than naively derived using dimensional analysis, i.e., $10^{18}$ GeV. In models with extra-dimensions \cite{ArkaniHamed:1998rs,Randall:1999ee}, the large extra-dimensional volume could explain why gravity appears weak from the four dimensional point of view, while in four  dimensional models with a large hidden sector \cite{Calmet:2008tn, Dvali:2001gx} the running of the Planck mass  lowers the scale of quantum gravity to the TeV region. It has recently been shown that in models with a large number of fields, unitarity of the S-matrix is violated well below the scale at which quantum gravity effects become strong \cite{Atkins:2010eq}.  In this paper we show that a similar problem appears in models with a large extra-dimensional volume designed to address the hierarchy problem. Models with low scale (i.e. 1 TeV) quantum gravity suffer from unitarity problems. These models are inconsistent as such and need to be embedded into non-local models of quantum gravity such as e.g. little string models \cite{Antoniadis:2001sw, Calmet:2007je} with low string scale in the TeV region below the Planck mass. 

In Section 2 of this paper, we reconsider four-dimensional models of low scale quantum gravity based on a renormalization of Newton's constant. Besides the large hidden sector scenario which has already been extensively studied \cite{Calmet:2008tn,Calmet:2009gn,Calmet:2008dg}, we show that it is also possible to modify the value of the energy scale at which gravity becomes strong using the non minimal coupling of a scalar field to the Ricci scalar.  In Section 3, we study the unitarity bounds on the number of particles and the non minimal coupling and discuss applications to different models ranging from models designed to address the hierarchy problem, grand unified theories and inflationary models. These bounds are derived from the gravitational scattering of the particles of the model. In Section 4, we extend our considerations to models with a large extra-dimensional volume and show that unitarity is violated below the scale at which gravity is expected to become strong, the reason for this problem is the large number of Kaluza-Klein modes of the graviton. Although the result is similar to that obtained for the four dimensional models, the physics is quite different. We consider the scattering of standard model particles through Kaluza-Klein copies of the graviton. Finally we conclude. Our main conclusion is that if these models are relevant to nature, their first signal at the Large Hadron Collider would not be of a gravitational nature such as graviton emission or small black holes, but rather linked to the mechanism which fixes the unitarity problem. It is possible to lower the scale of quantum gravity down to 14 TeV without a violation of unitarity below that scale if one considers a four dimensional model with a singlet scalar field that is strongly non minimally coupled to the Ricci scalar. Finally we also show that models of inflation where the Higgs boson of the standard model is the inflaton are inconsistent.

\section{The running of Newton's constant: non-minimal coupling}

We start by considering the usual four dimensional Einstein-Hilbert action coupled to real scalar fields, Weyl fermions and vector fields treating them as massless particles
\begin{eqnarray} \label{effac1}
S[g,\phi, \psi, A_\mu]&= -\int d^4x \sqrt{-\det(g)} & \left   (\frac{1}{16\pi G_N} R+  \frac{1}{2} g^{\mu\nu} \partial_\mu \phi 
 \partial_\nu \phi + \xi R \phi^2 + \right .  \\ && \nonumber \left . +  e \bar \psi i \gamma^\mu D_\mu \psi + \frac{1}{4} F_{\mu\nu}F^{\mu\nu} \right)
\end{eqnarray}
where $e$ is the tetrad, $D_\mu=\partial_\mu + w^{ab}_\mu \sigma_{ab}/2$ and $ w^{ab}_\mu$ is the spin connection which can be expressed in terms of the tetrad, finally $\xi$ is the non-minimal coupling.
 The running of the reduced Planck mass due to a non-minimally coupled real scalar field can be deduced from the running of Newton's constant see  \cite{Calmet:2008tn} and \cite{Larsen:1995ax,Vassilevich:1994cz,Kabat:1995eq}:
\begin{eqnarray}
\bar M(\mu)^2=\bar M(0)^2-\frac{1}{16 \pi^2} \left  (\frac{1}{6} N_l + 2 \xi N_\xi \right ) \mu^2
\end{eqnarray}
where $\mu$ is the renormalization scale and $N_l=N_S+N_F-4 N_V$ where $N_S$, $N_F$ and $N_V$ are respectively the numbers of real scalar fields, Weyl fermions and vector bosons in the model and $N_\xi$ is the number of real scalar fields in the model with a non-minimal coupling to gravity. Note that the conformal value of $\xi $ in our convention is $1/12$. The scale at which quantum gravitational effects become strong, $\mu_\star$, follows from the requirement that the reduced Planck mass at this scale $\mu_\star$ be comparable to the inverse of the size of the fluctuations of the geometry, in other words, $\bar M(\mu_\star)\sim \mu_\star$.  We find:
\begin{eqnarray}\label{mustar}
\mu_\star=\frac{\bar M(0)}{\sqrt{1+\frac{1}{16 \pi^2} \left  (\frac{1}{6} N_l + 2 \xi N_\xi \right )}}.
\end{eqnarray}
Clearly the energy scale at which quantum gravitational effects become relevant depends on the number of fields introduced in the theory and on the non-minimal coupling $\xi$. While minimally coupled spin 0 and spin $1/2$ fields lower $\mu_\star$, spin 1 increase the effective reduced Planck mass and non-minimally coupled scalar fields can increase or lower $\mu_\star$ depending on the algebraic sign of $\xi$. The contribution of the graviton is a $1/N$ effect and very small if $N$ is reasonably large. 

The contribution of non-minimally coupled scalar fields to the renormalization of $\mu_\star$ could have intriguing effects for grand unified models. It was shown in \cite{Calmet:2008df} that dimension five operators  previously considered in the literature  \cite{Rizzo:1984mk,Hill:1983xh,Shafi,Hall:1992kq,Vayonakis:1993nn,Datta:1995as,Dasgupta:1995js,Huitu:1999eh,Tobe:2003yj,Chakrabortty:2008zk}:
\begin{eqnarray}
\frac{c_i}{\mu_\star}H_i^{ab}G^a_{\mu\nu}G^{b\mu\nu}~,
\end{eqnarray}
where $G_{\mu\nu}$ is the gauge field strengths of the unified theory, $H_i$ are the Higgs multiplets  of the grand unified gauge group 
and $c_i$ are dimensionless Wilson coefficients, can be much more important than previously assumed because of the renormalization of the reduced Planck mass. This effect is particularly important if the grand unified theory contains Higgses coming in large representations of the unification group. Since these operators modify the unification condition, naive extrapolation using low energy data is not possible and unification models which apparently do not lead to satisfactory unification could actually fulfill the modified unification condition \cite{Calmet:2008df}.  Here we wish to point out that even in models with a small number of fields, the running of the reduced Planck mass can be sizable due to the non-minimal coupling of a real scalar field which transforms as a singlet under the unification group. This effect opens up the door to a very wide range of values for $\mu_\star$. Note that as we shall see below the coefficient of the non-minimal coupling between the Ricci scalar and the Higgses which transform in non trivial representations of the gauge group cannot be large because of unitarity issues.

The case of a large hidden sector composed of minimally coupled scalar fields and Weyl fermions has been studied in \cite{Calmet:2008tn}. It is shown that $10^{33}$ such fields would lead to a reduced Planck mass of 1 TeV. It was later shown that this scenario leads to a violation of unitarity below this scale \cite{Atkins:2010eq}. It is interesting to point out that the same effect, i.e. $\mu_\star= 1 \ {\rm TeV}$ could be reached by taking one single real scalar field with a large non-minimal coupling, i.e., $\xi=4.6 \times 10^{32}$. However as we shall see below, unitarity implies that this scalar field has to be extremely light and this scenario is ruled out by searches for deviations of the $1/r$ Newton potential. However, $\mu_\star$ could be as low as $14 \ {\rm TeV}$ without being in conflict with any current data.

Other extreme cases are of interest. It is possible to make gravity asymptotically free  by introducing $237$ vector bosons. In which case gravity never becomes strongly coupled and there is no minimal length in the model \cite{minlength}. We are assuming here that strong gravitational effects are not introduced by higher dimensional operators of the type $R^2$ etc. If a well defined effective theory exists for gravity, it would be surprising for these higher dimensional effects to reverse the effect obtained from the leading order term, i.e., the Ricci scalar.  The same effect can be obtained using a single non-minimally coupled scalar field with $\xi=-8\pi^2-1/12\sim -78.9$. We could also have used a combination of spin 1 and non-minimally coupled spin 0 fields.

These examples show that the energy scale at which gravitational quantum effects become relevant strongly depends on the particle content. We shall see in the next section that the requirement of unitarity imposes strong constraints on the content of particles coupled to linearized general relativity.

\section{Unitarity of linearized general relativity coupled to matter fields}

In \cite{Atkins:2010eq} we have introduced a criteria for the validity of linearized general relativity. We compared the scale at which unitarity is violated, $E_\star$, in the gravitational scattering of particles of spin 0, 1/2 and 1 to the scale at which quantum gravitational effects become strong i.e. $\mu_\star$. The tree level amplitudes had been obtained previously in \cite{Han:2004wt}. Using our criteria, we derived a bound on the particle content of a model coupled to linearized general relativity. We found  \cite{Atkins:2010eq}:
\begin{eqnarray} \label{bound1}
\frac{N_S}{3}+N_F+4 N_V \le  160 \pi
\end{eqnarray}
using the $J=2$ partial wave and 
\begin{eqnarray} \label{bound2}
(1+12 \xi)^2 N_S \le  96 \pi
\end{eqnarray}
using the $J=0$ partial wave. These bounds are obtained considering gravitational scattering of the type  $ 2 \phi_i \to 2 \phi_j$ with $i\ne j$ which are $s$-channel processes. Imposing different ingoing and outgoing particles ensures the absence of $t$ and $u$-channels. These bounds are thus valid for $N_i>1$.

A direct consequence of the second bound is a bound on the non-minimal coupling $\xi$:
\begin{eqnarray} \label{boundxi}
-\frac{ 4\sqrt{6 \pi N_\xi}  +N_\xi }{12 N_\xi} \le \xi \le \frac{ 4\sqrt{6 \pi N_\xi}  -N_\xi }{12 N_\xi}
\end{eqnarray}
where $N_\xi$ is the number of real scalar fields which couple to the Ricci scalar with the coupling $\xi$. This has direct implications for models of inflation where the Higgs boson of the standard model is strongly coupled to the Ricci scalar.
Indeed, in the standard model there is one Higgs doublet and hence four real scalars. Thus $N_\xi=4$ and we find  $-0.81\le \xi \le 0.64$ numerically. Note that the conformal value, $\xi=-1/12=-0.083$, is within this range and that in the limit $N_\xi \to \infty$, $\xi$ is forced to take the conformal value. If the model under consideration is to be valid up to the reduced Planck mass, the parameter $\xi$ needs to be rather small and is theoretically very tightly constrained. Clearly this casts some serious doubts on the validity of certain inflationary models such as, for example, the models proposed in \cite{Bezrukov:2007ep,DeSimone:2008ei} where the Higgs boson plays the role of the inflaton.  These models, although beautiful and minimalistic, require some new physics below the reduced Planck mass to fix the unitarity problem. Let us now consider these models of inflation where the Higgs doublet of the standard model plays the role of the inflaton \cite{Bezrukov:2007ep,DeSimone:2008ei} in more detail. Obtaining the correct number of e-foldings  requires $\xi \sim 10^4$. Unitarity is violated at  an energy 
\begin{eqnarray}
E_\star=\bar M \frac{4 \sqrt{6 \pi}}{1+12 \xi}.
\end{eqnarray}
For $\xi \sim 10^4$, one finds $E_\star=3.5 \times 10^{14}$ GeV. On the other hand, from equation (\ref{mustar}), we see that for large $\xi$, quantum gravity effects become strong at an energy scale
\begin{eqnarray}
\mu_\star \sim \frac{2 \sqrt{2} \pi \bar M(0)}{\sqrt{\xi N_\xi}}.
\end{eqnarray}
For $\xi=10^4$ and $N_\xi=4$ one finds  $\mu_\star \sim 1 \times 10^{17} \ \mbox{GeV}$. Clearly the model breaks down unless some new physics is introduced to restore unitarity. It is not trivial to see what kind of physics could be introduced to do so without putting the stability of the inflation potential into jeopardy. Our treatment of the problem which takes into account quantum gravitational effects is compatible with previous observations \cite{Atkins:2010eq} see also \cite{Burgess:2009ea,Barbon:2009ya,Burgess:2010zq,Hertzberg:2010dc}. It should be emphasized that the scheme used to derive the physical properties of the model cannot affect the outcome of the conclusions we are making.

In models with multiple scalar fields which have distinct non-minimal couplings we could also consider the process $S_A + S_A \to S_B + S_B$, where $S_A$ and $S_B$ are real scalar fields with non-minimal couplings $\xi_A$ and $\xi_B$ respectively. The bound we would obtain from this process would be
\begin{eqnarray}
(1+12\xi_A)(1+12\xi_B)\sqrt{N_A N_B} \le 96 \pi
\end{eqnarray}
where $N_A$ and $N_B$ are the number of scalar fields with couplings $\xi_A$ and $\xi_B$ respectively. One might think that by introducing a field with a coupling $\xi \sim 10^{-4}$ might fix the above unitarity problem with the Higgs, but it should be remembered that the strongest bound will still come from (\ref{bound2}).

We emphasize that the bound on $\xi$ is valid for $N_\xi>1$ since we want to avoid the discussion of the $t$ and $u$ channels. It turns out that for $N_\xi=1$ a cancellation of the terms proportional to $\xi$ growing with energy happens between the $s$, $t$ and $u$ channels. The amplitude for $S+S\to S+S$ where $S$ is a real scalar field is given by \cite{Huggins:1987ea}
\begin{eqnarray} 
A_\xi&=&\frac{-2}{\bar M_P^2}\left( m^4(s^{-1}+t^{-1}+u^{-1})+\frac{(2m^2-s)(2m^2-t)}{2u}+\frac{(2m^2-s)(2m^2-u)}{2t}
+ \right . \\  \nonumber&& \left . +\frac{(2m^2-s)(2m^2-u)}{2t}+2m^2\xi(6 \xi -5)\right)
\end{eqnarray}
where $m$ is the mass of the scalar field. We are interested in the high energy limit where all invariants $s$, $t$ and $u$ are large i.e.  $s \sim |t| \sim |u|> s_{\rm min}$ there is no term growing with energy proportional to $\xi$. However we can use the $J=0$ partial wave bound to set a limit on the mass of the scalar field.  In the high energy limit one finds
\begin{eqnarray} 
A_\xi&\sim&\frac{-2}{\bar M_P^2}\left(\frac{3}{2} s +2m^2\xi(6 \xi -5)\right)
\end{eqnarray}
Setting $\bar M_P \sim \sqrt{s} \sim E_\star \sim \mu_\star$, we find:
\begin{eqnarray} \label{boundmass}
\left | \frac{3}{2} + \frac{2 m^2}{\bar M_P^2} \xi (6 \xi -5) \right | \le 4 \pi
\end{eqnarray}
using the $J=0$ partial wave bound. For the large $\xi \sim 10^{32}$ model discussed above, we find $m<2 \times 10^{-14} \ {\rm GeV}$. This bound is analogous to the bound on the Higgs boson's mass in the standard model. Such a large $\xi$ corresponds to interactions with a strength comparable to that of gravity. A very light 5th force carrier with a mass low enough to satisfy our bound is clearly ruled out by probes of the Newtonian $1/r$ potential. The non-minimal coupling cannot be used to lower the effective Planck mass down to  1 TeV.  However, we shall see that it could be lowered to $14 \ {\rm TeV}$.

Let us now reconsider the extreme cases discussed in the previous section in the light of the bounds (\ref{bound1}) and  (\ref{bound2}).  Clearly the model with $10^{33}$ scalars and or fermions has issues with unitarity below the scale at which quantum gravitational effects become strong. As mentioned above, the large $\xi \sim 10^{32}$ is ruled out by observations. The problem with unitarity could be solved by embedding this model into a non-local theory of gravity such as string theory. If the string scale is lower than the reduced Planck mass, then non-local effects due to the stringy nature of the model could affect our conclusions and restore unitarity up to the reduced Planck mass. It is thus difficult to lower the reduced Planck mass using the renormalization of Newton's constant. Could we, up to the caveat mentioned in the previous section,  increase the reduced Planck mass and make gravity asymptotically free? We mentioned that 237 vector bosons could make gravity asymptotically free, however this is in clear violation of the bound  (\ref{bound1}).
Similarly, the other scenario mentioned to make gravity asymptotically  free by taking $\xi= -78.9$ is ruled out by our bound  (\ref{bound2}) if there is more than one real scalar field non-minimally coupled to the Ricci scalar. A single real scalar field non-minimally coupled to the Ricci scalar with a $\xi= -78.9$ could however make gravity asymptotically free, assuming the caveat mentioned above is fulfilled. In that world there would be no fundamental length, as small quantum black holes would never form  \cite{minlength}.

It is easy to convince oneself that the lowest value for $\mu_\star$ one can reach  without violating unitarity below that scale is obtained by taking $N_V=0$, $\xi=0$, $N_S=301$ and $N_F=401$. These parameters correspond to $\mu_{\star, \ {\rm min}}=0.76 \bar M_P$.  This bound is valid for $N_S>1$, but can be circumvented if we consider a single real scalar field non-minimally coupled to the Ricci scalar.  We could choose a mass of $10^{-3} \ {\rm eV}$ without being in conflict with experiments probing for modification of Newton's $1/r$ potential as we shall see that the interaction of this field with matter will be slightly weaker than that of gravity. The unitarity bound on the real scalar field mass  (\ref{boundmass}) allows us to pick $\xi=2.3 \times 10^{30}$ without violating unitarity. This corresponds to a $\mu_\star=14  \ {\rm TeV}$ which is the reduced Planck mass in that model. This simple model is the only one able to lower the scale of quantum gravity dramatically without suffering from unitarity problems. The strength of the interaction of the real scalar with matter is just slightly weaker than that of gravity.  One may worry that such a large non-minimal coupling could lead to a breakdown of the perturbative expansion well below $\mu_\star$. However, one can easily see that the contribution at one loop is of the order $\xi/M_P^2$, the contribution proportional to $\xi^2/M_P^2$ vanishes  by the equation of motion. Higher orders will be suppressed by powers of $1/(16 \pi^2)$.

So far we have considered the unitarity bound due to the matter content of the theory.  One may worry that unitarity violation occurs in linearized general relativity without matter. To check this we consider graviton gravitational scattering. Graviton gravitational scattering has been considered before in the literature \cite{DeWitt:1967uc,Berends:1974gk,Grisaru:1975bx}. Using the helicity amplitudes for graviton graviton scattering, one finds:
\begin{equation}\label{channel 1}
A_{++;++}=\kappa^2 \frac{s^4}{4stu}
\end{equation}
\begin{equation}\label{channel 2}
A_{+-;+-}=\kappa^2 \frac{u^4}{4stu}
\end{equation}
\begin{equation}\label{channel3 }
A_{-+;+-}=\kappa^2 \frac{t^4}{4stu}
\end{equation}
where $\kappa^2=32\pi G = 4/\bar M_p^2$. The $s$-channel is vanishing and we thus have to rely on the $t$ and $u$ channels which are more troublesome. Strictly speaking, the partial wave expansions are not defined for these amplitudes since they have poles at $\theta =\pm \pi$ in the denominator. However we are interested in studying the high energy limit of these amplitudes and in particular the regime where $s \sim |t| \sim |u| >s_{\rm min}$, i.e., the regime where all invariants are large. These amplitudes thus behave as $\sim s/{\bar M_P^2}$ in the high energy limit. We can obtain a unitarity bound from the $J=0$ partial wave and find that
\begin{eqnarray}
\frac{\sqrt{s}}{\bar M_P} \le \sqrt{ 8 \pi}.
\end{eqnarray}
In other words, unitarity violation occurs above the reduced Planck mass and strong gravitational effects are expected to cure this problem.

We have shown in this section that the constraint of unitarity implies that models with a large hidden sector designed to bring down the reduced Planck mass to the TeV region will need to be modified to fix the unitarity issue. The simplest possibility seems to be to embed these models \cite{Calmet:2008tn, Dvali:2001gx}  into non-local theories in which case the experimental signatures of these models will be different than those studied so far in the literature, namely graviton \cite{Calmet:2009gn} and black hole productions \cite{Calmet:2008dg}. One naively expects to first find some signal of the non-local nature of these models in the form of some extension in space of the particles of the standard model. If the UV completion of these models is little string theory, one first expects to discover non gravitational stringy effects, see e.g. \cite{Anchordoqui:2009mm}. The only way to bring down the reduced Planck mass within reach of the Large Hadron Collider is to use a single real scalar field with a large non-minimal coupling to the Ricci scalar. The lowest value of the reduced Planck mass one can obtain given the constraint from experiments looking for deviations of Newton's $1/r$ potential is $14 \ {\rm TeV}$.

In the next section we shall consider scenarios with more than four-dimensions designed to lower the reduced Planck mass in the TeV region. We shall show that the same conclusion can be reached: strong gravitational effects will not be the first signals  at the Large Hadron Collider of brane world models even if one of them is realized in nature.

\section{Unitarity of models with large extra-dimensions}

Models with large extra-dimesnions have become extremely popular over the last decade, in large part due to them providing a geometric reformulation of the gauge hierarchy problem. In general these models allow gravity to propagate in the bulk while matter fields are confined to a 3-brane. The extra-dimensions are compact, and from a four dimensional effective field theory perspective this manifests itself via the presence of a tower of massive Kaluza Klein (KK) gravitons.

We consider s-channel scattering of matter particles via exchange of KK gravitons. The process is identical to that considered in \cite{Atkins:2010eq} but with massive gravitons appearing in the propagator. The propagator for a massive graviton, even in the massless limit, differs from that of a massless graviton by the well known van Dam-Veltman-Zakharov discontinuity \cite{vanDam:1970vg}. The amplitudes for s-channel scattering via KK gravitons in the massless limit are presented in table (1). The partial wave amplitudes, $a_J$, can be determined using ${\cal A} =16 \pi \sum_J (2J+1) a_J d^J_{\mu,\mu^\prime}$. Comparing the amplitudes with those calculated for a massless graviton in \cite{Atkins:2010eq} we can immediately see that while the $J=2$ partial waves are the same in both cases, the $J=0$ partial wave is absent when massive gravitons are exchanged. The results are not the same because of the presence of an extra scalar degree of freedom for the massive gravitons.

\begin{table*}[tbh]
\resizebox{\textwidth}{!}{
%\begin{center}
\begin{tabular}{|c|c|c|c|c|c|}  
\hline
 $\to$ & $s' s'$ & $\psi'_+\bar \psi'_- $ & $ \psi'_-\bar \psi'_+ $ & $V'_+ V'_-$ & $V'_- V'_+$ \\
\hline $s s$  & $-2/3\pi G_N s\ d^2_{0,0} $ & $ -2\pi G_N s \sqrt{1/3}\ d^2_{0,1}$
& $ -2\pi G_N s \sqrt{1/3}\ d^2_{0,-1}$  & $-4\pi G_N s \sqrt{1/3}\ d^2_{0,2} $  & $ -4\pi G_N s\sqrt{1/3}\ d^2_{0,-2} $\\
\hline $\psi_+\bar \psi_- $ &
                $ -2\pi G_N s \sqrt{1/3}\ d^2_{1,0}$ & $ -2\pi G_N s\ d^2_{1,1} $
& $  -2\pi G_N s\ d^2_{1,-1} $ &  $-4\pi G_N s\ d^2_{1,2}$ & $-4\pi G_N s \  d^2_{1,-2}$ \\
\hline $\psi_-\bar \psi_+ $ &
                $-2\pi G_N s \sqrt{1/3}\ d^2_{-1,0}$ & $ -2\pi G_N s\ d^2_{-1,1} $
& $ -2\pi G_N s\ d^2_{-1,-1} $ &   $ -4\pi G_N s\ d^2_{-1,2}$ & $ -4\pi G_N s \ d^2_{-1,-2}$ \\
\hline $V_+ V_- $ &  $ -4\pi G_N s \sqrt{1/3}\ d^2_{2,0}$ &  $ -4\pi G_N s\
d^2_{2,1}$ &$
 -4\pi G_N s\ d^2_{2,-1}  $ & $-8\pi G_N s\ d^2_{2,2}$ & $-8\pi G_N s\ d^2_{2,-2}$ \\
\hline $V_- V_+  $ &  $-4\pi G_N s\sqrt{1/3}\ d^2_{-2,0}$ &  $ -4\pi G_N s\
d^2_{-2,1}$ &$
 -4\pi G_N s\ d^2_{-2,-1}  $ & $-8\pi G_N s\ d^2_{-2,2}$ & $-8\pi G_N s\ d^2_{-2,-2}$ \\
\hline  
\end{tabular}}
\caption{Scattering amplitudes for scalars, fermions, and vector
bosons via s-channel KK graviton exchange in terms of the Wigner
$d$ functions \cite{Amsler:2008zzb} in the massless limit.  $G_N=1/M_P^2$ is Newton's constant and $s=E^2_{\rm{CM}}$ is the center of mass energy squared. We have used the helicity basis as in  \cite{Han:2004wt}}  \label{t1}
%\end{center}
\end{table*}

Each partial wave is subject to the unitarity bound  $| \mbox{Re} \ a_J|\le1/2$. Considering the $J=2$ partial wave for the scattering of a superposition of states, $|\sqrt{1/3} \sum s  s + \sum \psi_- \bar \psi_++ 2 \sum V V \rangle$, we aquire the same unitarity bound as that found in \cite{Atkins:2010eq}:
\begin{equation}\label{oneKKbound}
|a_2|=\frac{1}{320 \pi} \frac{s}{\bar M_P^2}N \le\frac12
\end{equation}
where $N=1/3 N_S + N_F + 4N_V$.

Each amplitude in table (1) can however occur via exchange of any of a very large number of KK gravitons. If we consider the total amplitude for each process to occur via exchange of all possible KK gravitons the entries in table (1) will get multiplied by a global factor of the number of KK modes. Including $N_{KK}$ gravitons, we find the bound becomes
\begin{equation}\label{multiKKbound}
|a_2|=\frac{1}{320 \pi} \frac{s}{\bar M_P^2}N_{KK} N \le\frac12 .
\end{equation}

Now let us consider the model \cite{ArkaniHamed:1998rs} with $n$ extra-dimensions and the extra-dimensional fundamental Planck scale, $M_D$, given by $\bar M_P^2 =  R^n M_D^{2+n}$, where the extra-dimensions are compactified on a $n$-dimensional torus with common radius $R$. We take the fundamental Planck mass $M_D = 1$ TeV, as is usually done in order to address the hierachy problem. $M_D$ now defines the scale at which gravity becomes strong and where we expect our effective theory to break down, hence we should properly only include KK modes with masses below this scale. We find the number of KK gravitons with masses below 1 TeV is $N_{KK} \sim 10^{32}$. Inserting this into (\ref{multiKKbound}), and considering scattering of standard model particles, $N=N_{SM}=283/3$, we find that at $\sqrt{s}=1$  TeV, $|a_2|\sim 1.6$ and we see that unitarity is clearly violated below 1 TeV. The actual energy at which unitarity is violated is $E_\star= 0.55$ TeV. We note that in the Randall Sundrum model \cite{Randall:1999ee}, there are typically also a very large number of KK modes of the graviton, however the mass gap with the zero mode is larger.

The above calculation is of course only approximate as we have taken the massless limit for the KK gravitons. Since we are including all KK gravitons with masses up to 1 TeV we should properly take their masses into account. For exchange of a KK graviton with mass $m_i$, each of the amplitudes in table (1) should be multiplied by $s/(s-m_i^2)$. Because of the very small spacing of the KK masses in models with large extra-dimensions, we can approximate summing up all of the amplitudes for each $m_i$ by an integral. The number of modes with masses between $m$ and $m+dm$ is given by
\begin{equation}\label{dN}
dN= S_{n-1} m^{n-1} R^n dm 
\end{equation}
where $S_{n-1}=2 \pi^{n/2}/ \Gamma(n/2)$ is the surface of a unit-radius sphere in $n$ dimensions. Summing all the modes with masses $m_i \le E$, we find
\begin{equation}\label{sumint}
\sum_i \frac{1}{s-m_i^2} \approx \int_0^{E}\frac{1}{s-m^2}S_{n-1} m^{n-1} R^n dm .
\end{equation}
for $E<\sqrt{s}$. The integral clearly diverges when $E$ approaches $\sqrt{s}$ as the KK gravitons become on shell and so we should not include modes with masses too close to this energy. However, considering the model as above and only including KK modes with masses up to $95\%$ of $\sqrt{s}= 1$ TeV we find $|a_2|= 1.0, 1.3, 1.5, 1.5$ and $1.4$ for $n= 3,4,5,6$ and $7$ respectively and unitarity is again clearly violated below this energy  ($n=2$ and $n=3$ with $M_D=1 \ {\rm TeV}$ are ruled out from astrophysical bounds \cite{Hannestad:2001jv,Hannestad:2001xi} while $n=1$ and $M_D=1\ {\rm TeV}$  implying deviations from Newtonian gravity over solar system distances is clearly ruled out). Table (\ref{t2}) lists the actual value, $E_\star$ at which unitarity is violated depending on the number of KK modes we include in the calculation. We include KK modes with masses $m < m_{\rm{max}}$ where $m_{\rm{max}}<E_\star$. It is clear that even when a conservative number of KK modes are included, unitarity  is still violated below the energy at which gravity is supposed to become strong.

\begin{table*}[tbh]
\begin{center}
  \begin{tabular}{ | c  || c | c | c | c | c | c | c| }
    \hline
$m_{\rm{max}}$		& $n=1$	& $n=2$	&  $n=3$&  $n=4$&  $n=5$&  $n=6$ & $n=7$ \\ \hline
0.9 $E_\star$		& 1.2	& 1.0	&  0.94	&  0.93	&  0.93	&  0.95  & 0.97	 \\ \hline
0.95 $E_\star$		& 1.1	& 0.92	&  0.86	&  0.85	&  0.86	&  0.87	 & 0.89	 \\ \hline
0.99 $E_\star$		& 1.0	& 0.81	&  0.76	&  0.75	&  0.76	&  0.78	 & 0.80	 \\ \hline
0.999 $E_\star$		& 0.88	& 0.72	&  0.69	&  0.69	&  0.70	&  0.72	 & 0.75	 \\ \hline 
 \end{tabular}
\caption{Values of the energy, $E_\star$, in TeV at which unitarity is violated as a function of the number of KK modes considered. We include all KK modes with masses $m \le m_{\rm{max}}$. The energy is also dependent on the number of extra-dimensions, $n$.   } 
\label{t2}
\end{center}
\end{table*}

Table (\ref{t2}) clearly shows that the closer we pick the cutoff to $M_D$ the more serious the unitarity constraint becomes. However it is not clear how close to $E_\star$ we can take $m_{\rm{max}}$. To get a better feel for this we can introduce a width for the KK gravitons $\Gamma(m) \sim \frac{m^3}{\bar M_P^2}$. We can now sum up all the KK modes with masses $m \le E_\star$ using
\begin{equation}\label{sumintEstar}
\int_0^{E_\star}\frac{1}{E_\star-m^2 + i m \Gamma(m)}S_{n-1} m^{n-1} R^n dm ,
\end{equation}
and we find that unitarity is violated at $E_\star = 0.47, 0.49, 0.53, 0.56$ and $0.59$ TeV for $n=3, 4, 5, 6$ and $7$ respectively.

Increasing the extra-dimensional Planck mass, $M_D$, does not improve the situation. While we find much less KK modes below $M_D$, the partial wave is proportionally increased by considering $s=M_D^2$. We do not find any scale for $M_D$ for which  this unitarity problem does not exist.

We believe this violation of unitarity in large extra-dimensional models should be taken seriously. It clearly shows that new physics must appear below 1 TeV to fix the unitarity issue. We would therefore expect to see signatures of this new physics at the LHC experiment well before we would see other signatures of these models such as small black holes. Finally we note that first indications of a violation of unitarity below the scale $M_D$ had been reported in e.g. \cite{Kachelriess:2000cb}, but the channel we considered which leads to a tight constraint had not been considered previously. Unitarity constraints have also previously been considered for the Randall Sundrum model \cite{Grzadkowski:2007zz}.

\section{Conclusions}

 We have shown that models with low scale (i.e. 1 TeV) quantum gravity suffer from unitarity problems below this energy scale. This applies to four dimensional models as well as to models with a  large extra-dimensional volume designed to address the hierarchy problem.  These models are inconsistent as such and need to be embedded into non-local models of quantum gravity such as  string models with a low string scale in the TeV region below the scale at which gravity becomes strong. 

We have reconsidered four-dimensional models of low scale quantum gravity based on a renormalization of Newton's constant. Besides the large hidden sector scenario which has been already study extensively  \cite{Calmet:2008tn,Calmet:2009gn,Calmet:2008dg}, we have shown that it is also possible to modify the value of the energy scale at which gravity becomes strong using the non minimal coupling of a scalar field to the Ricci scalar.  We studied the unitarity bounds on the number of particles and the non minimal coupling and discussed applications to different models ranging from models designed to address the hierarchy problem, grand unified theories to inflationary models. 

We then turned our considerations to models with a large extra-dimensional volume and show that unitarity is violated below the scale at which gravity is expected to become strong, the reason for this problem is the large number of Kaluza-Klein modes of the graviton.    Although the result is similar to that obtained for the four dimensional models, the physics is quite different. We considered the scattering of standard model particles through Kaluza-Klein copies of the graviton.

Our main conclusion is that if these models are relevant to nature, their first signal at the Large Hadron Collider would not be of a gravitational nature such as graviton emission or small black holes, but rather linked to the mechanism which fixes the unitarity problem. It is possible to lower the scale of quantum gravity down to 14 TeV without a violation of unitarity below that scale if one considers a four dimensional model with a singlet scalar field that is strongly non minimally coupled to the Ricci scalar. Finally we have shown that models of inflation where the Higgs boson of the standard model is the inflaton are inconsistent.

\bigskip

{\it Acknowledgments:} 
We would like to thank M. Fairbairn and  F. Maltoni for a helpful discussion. This work is supported in part by the European Cooperation in Science and Technology (COST) action MP0905 "Black Holes in a Violent Universe". 

%\newpage

%%%%%%%%%%%%%%%%%%%%%%%%%%%%%%%%%%%%%%%%%%%%%%%%%%%%%%%%%%%%%%%%%
%%%
%%%                     BIBLIOGRAPHY
%%%
%%%%%%%%%%%%%%%%%%%%%%%%%%%%%%%%%%%%%%%%%%%%%%%%%%%%%%%%%%%%%%%%%

\bigskip

%\newpage
%\vskip .75 in


\baselineskip=1.6pt
\begin{thebibliography}{99}

%\cite{ArkaniHamed:1998rs}
\bibitem{ArkaniHamed:1998rs}
  N.~Arkani-Hamed, S.~Dimopoulos and G.~R.~Dvali,
  %``The hierarchy problem and new dimensions at a millimeter,''
  Phys.\ Lett.\  B {\bf 429}, 263 (1998)
  [arXiv:hep-ph/9803315];
  %%CITATION = PHLTA,B429,263;%%
%\cite{Antoniadis:1998ig}
%\bibitem{Antoniadis:1998ig}
  I.~Antoniadis, N.~Arkani-Hamed, S.~Dimopoulos and G.~R.~Dvali,
  %``New dimensions at a millimeter to a Fermi and superstrings at a TeV,''
  Phys.\ Lett.\  B {\bf 436}, 257 (1998)
  [arXiv:hep-ph/9804398].
  %%CITATION = PHLTA,B436,257;%%


%\cite{Randall:1999ee}
\bibitem{Randall:1999ee}
  L.~Randall and R.~Sundrum,
  %``A large mass hierarchy from a small extra dimension,''
  Phys.\ Rev.\ Lett.\  {\bf 83}, 3370 (1999)
  [arXiv:hep-ph/9905221].
  %%CITATION = PRLTA,83,3370;%%


%\cite{Calmet:2008tn}
\bibitem{Calmet:2008tn}
  X.~Calmet, S.~D.~H.~Hsu and D.~Reeb,
  %``Quantum gravity at a TeV and the renormalization of Newton's constant,''
  Phys.\ Rev.\  D {\bf 77}, 125015 (2008)
  [arXiv:0803.1836 [hep-th]];
  %%CITATION = PHRVA,D77,125015;%%
%\cite{Calmet:2010qq}
%\bibitem{Calmet:2010qq}
  X.~Calmet,
  %``Renormalization of Newton's constant and Particle Physics,''
  arXiv:1002.0473 [hep-ph], to appear in the proceedings of  the 12th Marcel Grossmann Meeting.
  %%CITATION = ARXIV:1002.0473;%%
  
  
%\cite{Dvali:2001gx}
\bibitem{Dvali:2001gx}
  G.~R.~Dvali, G.~Gabadadze, M.~Kolanovic and F.~Nitti,
  %``Scales of gravity,''
  Phys.\ Rev.\  D {\bf 65}, 024031 (2002)
  [arXiv:hep-th/0106058].
  %%CITATION = PHRVA,D65,024031;%%
  
  %\cite{Atkins:2010eq}
\bibitem{Atkins:2010eq}
  M.~Atkins and X.~Calmet,
  %``On the unitarity of linearized general relativity coupled to matter,''
  [arXiv:1002.0003 [hep-th]].
  %%CITATION = ARXIV:1002.0003;%%

  
  %\cite{Antoniadis:2001sw}
\bibitem{Antoniadis:2001sw}
  I.~Antoniadis, S.~Dimopoulos and A.~Giveon,
  %``Little string theory at a TeV,''
  JHEP {\bf 0105}, 055 (2001)
  [arXiv:hep-th/0103033].
  %%CITATION = JHEPA,0105,055;%%

%\cite{Calmet:2007je}
\bibitem{Calmet:2007je}
  X.~Calmet and S.~D.~H.~Hsu,
  %``TeV gravity in four dimensions?,''
  Phys.\ Lett.\  B {\bf 663}, 95 (2008)
  [arXiv:0711.2306 [hep-ph]].
  %%CITATION = PHLTA,B663,95;%%


% %\cite{Calmet:2009gn,Calmet:2008dg}


%\cite{Calmet:2009gn}
\bibitem{Calmet:2009gn}
  X.~Calmet and P.~de Aquino,
  %``Quantum Gravity at the LHC,''
  arXiv:0906.0363 [hep-ph];
  %%CITATION = ARXIV:0906.0363;%%
%\cite{Calmet:2009yw}
%\bibitem{Calmet:2009yw}
  X.~Calmet, P.~de Aquino and T.~G.~Rizzo,
  %``Massless versus Kaluza-Klein Gravitons at the LHC,''
  Phys.\ Lett.\  B {\bf 682}, 446 (2010)
  [arXiv:0910.1535 [hep-ph]].
  %%CITATION = PHLTA,B682,446;%%
  
%\cite{Calmet:2008dg}
\bibitem{Calmet:2008dg}
  X.~Calmet, W.~Gong and S.~D.~H.~Hsu,
  %``Colorful quantum black holes at the LHC,''
  Phys.\ Lett.\  B {\bf 668}, 20 (2008)
  [arXiv:0806.4605 [hep-ph]];
  %%CITATION = PHLTA,B668,20;%%
  %\cite{Calmet:2008rv}
%\bibitem{Calmet:2008rv}
  X.~Calmet and M.~Feliciangeli,
  %``Bound on four-dimensional Planck mass,''
  Phys.\ Rev.\  D {\bf 78}, 067702 (2008)
  [arXiv:0806.4304 [hep-ph]].
  %%CITATION = PHRVA,D78,067702;%%
  
  
%\cite{Larsen:1995ax,Vassilevich:1994cz,Kabat:1995eq}

%\cite{Larsen:1995ax}
\bibitem{Larsen:1995ax}
  F.~Larsen and F.~Wilczek,
  %``Renormalization of black hole entropy and of the gravitational coupling
  %constant,''
  Nucl.\ Phys.\  B {\bf 458}, 249 (1996)
  [arXiv:hep-th/9506066].
  %%CITATION = NUPHA,B458,249;%%
  
%\cite{Vassilevich:1994cz}
\bibitem{Vassilevich:1994cz}
  D.~V.~Vassilevich,
  %``QED on curved background and on manifolds with boundaries: Unitarity versus
  %covariance,''
  Phys.\ Rev.\  D {\bf 52}, 999 (1995)
  [arXiv:gr-qc/9411036].
  %%CITATION = PHRVA,D52,999;%%
  
%\cite{Kabat:1995eq}
\bibitem{Kabat:1995eq}
  D.~N.~Kabat,
  %``Black hole entropy and entropy of entanglement,''
  Nucl.\ Phys.\  B {\bf 453}, 281 (1995)
  [arXiv:hep-th/9503016].
  %%CITATION = NUPHA,B453,281;%%





%\cite{Calmet:2008df}
\bibitem{Calmet:2008df}
  X.~Calmet, S.~D.~H.~Hsu and D.~Reeb,
  %``Grand unification and enhanced quantum gravitational effects,''
  Phys.\ Rev.\ Lett.\  {\bf 101}, 171802 (2008)
  [arXiv:0805.0145 [hep-ph]];
  %%CITATION = PRLTA,101,171802;%%
%\cite{Calmet:2008er}
%\bibitem{Calmet:2008er}
 % X.~Calmet, S.~D.~H.~Hsu and D.~Reeb,
  %``Quantum Gravitational Effects and Grand Unification,''
  AIP Conf.\ Proc.\  {\bf 1078}, 432 (2009)
  [arXiv:0809.3953 [hep-ph]].;
  %%CITATION = APCPC,1078,432;%%
%\cite{Calmet:2009hp}
%\bibitem{Calmet:2009hp}
%  X.~Calmet, S.~D.~H.~Hsu and D.~Reeb,
  %``Grand unification through gravitational effects,''
  Phys.\ Rev.\  D {\bf 81}, 035007 (2010)
  [arXiv:0911.0415 [hep-ph]].
  %%CITATION = PHRVA,D81,035007;%%

\bibitem{Hill:1983xh}
  C.~T.~Hill,
  %``Are There Significant Gravitational Corrections To The Unification Scale?,''
  Phys.\ Lett.\  B {\bf 135}, 47 (1984).

\bibitem{Shafi}
  Q.~Shafi and C.~Wetterich, Phys.\ Rev.\ Lett.\  {\bf 52}, 875 (1984).


\bibitem{Rizzo:1984mk}
  T.~G.~Rizzo,
  %``Gravitational Corrections To The Unification Scale In SO(10) With A Low
  %Right-Handed Mass Scale,''
  Phys.\ Lett.\  B {\bf 142}, 163 (1984).

\bibitem{Hall:1992kq}
  L.~J.~Hall and U.~Sarid,
  %``Gravitational smearing of minimal supersymmetric unification predictions,''
  Phys.\ Rev.\ Lett.\  {\bf 70}, 2673 (1993)
  [arXiv:hep-ph/9210240].

\bibitem{Vayonakis:1993nn}
  A.~Vayonakis,
  %``Planck Scale corrections to gauge coupling unification,''
  Phys.\ Lett.\  B {\bf 307}, 318 (1993).

\bibitem{Datta:1995as}
  A.~Datta, S.~Pakvasa and U.~Sarkar,
  %``Gravitational uncertainties from dimension-six operators on supersymmetric GUT predictions,''
  Phys.\ Rev.\  D {\bf 52}, 550 (1995)
  [arXiv:hep-ph/9403360].

\bibitem{Dasgupta:1995js}
  T.~Dasgupta, P.~Mamales and P.~Nath,
  %``Effects of gravitational smearing on predictions of supergravity grand unification,''
  Phys.\ Rev.\  D {\bf 52}, 5366 (1995)
  [arXiv:hep-ph/9501325].

\bibitem{Huitu:1999eh}
  K.~Huitu, Y.~Kawamura, T.~Kobayashi and K.~Puolamaki,
  %``Generic gravitational corrections to gauge couplings in SUSY SU(5)  GUTs,''
  Phys.\ Lett.\  B {\bf 468}, 111 (1999)
  [arXiv:hep-ph/9909227].

\bibitem{Tobe:2003yj}
  K.~Tobe and J.~D.~Wells,
  %``Gravity-assisted exact unification in minimal supersymmetric SU(5) and its gaugino mass spectrum,''
  Phys.\ Lett.\  B {\bf 588}, 99 (2004)
  [arXiv:hep-ph/0312159].

\bibitem{Chakrabortty:2008zk}
  J.~Chakrabortty and A.~Raychaudhuri,
  %``A note on dimension-5 operators in GUTs and their impact,''
  Phys.\ Lett.\  B {\bf 673}, 57 (2009)
  [arXiv:0812.2783 [hep-ph]].


  





\bibitem{minlength}
%\cite{Calmet:2004mp}
%\bibitem{Calmet:2004mp}
  X.~Calmet, M.~Graesser and S.~D.~H.~Hsu,
  %``Minimum length from quantum mechanics and general relativity,''
  Phys.\ Rev.\ Lett.\  {\bf 93}, 211101 (2004)
  [arXiv:hep-th/0405033];
  %%CITATION = PRLTA,93,211101;%%
%\cite{Calmet:2005mh}
%\bibitem{Calmet:2005mh}
 % X.~Calmet, M.~Graesser and S.~D.~H.~Hsu,
  %``Minimum length from first principles,''
  Int.\ J.\ Mod.\ Phys.\  D {\bf 14}, 2195 (2005)
  [arXiv:hep-th/0505144];
  %%CITATION = IMPAE,D14,2195;%%
%\cite{Calmet:2007vb}
%\bibitem{Calmet:2007vb}
  X.~Calmet,
  %``On the precision of a length measurement,''
  Eur.\ Phys.\ J.\  C {\bf 54}, 501 (2008)
  [arXiv:hep-th/0701073].
  %%CITATION = EPHJA,C54,501;%%

%\cite{Han:2004wt}
\bibitem{Han:2004wt}
 T.~Han and S.~Willenbrock,
  %``Scale of quantum gravity,''
  Phys.\ Lett.\  B {\bf 616}, 215 (2005)
  [arXiv:hep-ph/0404182].
  %%CITATION = PHLTA,B616,215;%%


%\cite{Bezrukov:2007ep}
\bibitem{Bezrukov:2007ep}
  F.~L.~Bezrukov and M.~Shaposhnikov,
  %``The standard model Higgs boson as the inflaton,''
  Phys.\ Lett.\  B {\bf 659}, 703 (2008)
  [arXiv:0710.3755 [hep-th]].
  %%CITATION = PHLTA,B659,703;%%

%\cite{DeSimone:2008ei}
\bibitem{DeSimone:2008ei}
  A.~De Simone, M.~P.~Hertzberg and F.~Wilczek,
  %``Running Inflation in the standard model,''
  Phys.\ Lett.\  B {\bf 678}, 1 (2009)
  [arXiv:0812.4946 [hep-ph]].
  %%CITATION = PHLTA,B678,1;%%


  %\cite{Burgess:2009ea}
\bibitem{Burgess:2009ea}
  C.~P.~Burgess, H.~M.~Lee and M.~Trott,
  %``Power-counting and the Validity of the Classical Approximation During
  %Inflation,''
  JHEP {\bf 0909}, 103 (2009)
  [arXiv:0902.4465 [hep-ph]].
  %%CITATION = JHEPA,0909,103;%%


%\cite{Barbon:2009ya}
\bibitem{Barbon:2009ya}
  J.~L.~F.~Barbon and J.~R.~Espinosa,
  %``On the Naturalness of Higgs Inflation,''
  Phys.\ Rev.\  D {\bf 79}, 081302 (2009)
  [arXiv:0903.0355 [hep-ph]].
  %%CITATION = PHRVA,D79,081302;%%

%\cite{Burgess:2009ea,Barbon:2009ya,Burgess:2010zq,Barbon:2009ya}


%\cite{Burgess:2010zq}
\bibitem{Burgess:2010zq}
  C.~P.~Burgess, H.~M.~Lee and M.~Trott,
  %``Comment on Higgs Inflation and Naturalness,''
  arXiv:1002.2730 [Unknown].
  %%CITATION = ARXIV:1002.2730;%%

%\cite{Hertzberg:2010dc}
\bibitem{Hertzberg:2010dc}
  M.~P.~Hertzberg,
  %``On Inflation with Non-minimal Coupling,''
  arXiv:1002.2995 [Unknown].
  %%CITATION = ARXIV:1002.2995;%%


%\cite{Huggins:1987ea}
\bibitem{Huggins:1987ea}
  S.~R.~Huggins and D.~J.~Toms,
  %``One Graviton Exchange Interaction Of Nonminimally Coupled Scalar Fields,''
  Class.\ Quant.\ Grav.\  {\bf 4}, 1509 (1987).
  %%CITATION = CQGRD,4,1509;%%


%\cite{DeWitt:1967uc}
\bibitem{DeWitt:1967uc}
  B.~S.~DeWitt,
  %``Quantum theory of gravity. III. Applications of the covariant theory,''
  Phys.\ Rev.\  {\bf 162} (1967) 1239.
  %%CITATION = PHRVA,162,1239;%%



%\cite{Berends:1974gk}
\bibitem{Berends:1974gk}
  F.~A.~Berends and R.~Gastmans,
  %``On The High-Energy Behavior In Quantum Gravity,''
  Nucl.\ Phys.\  B {\bf 88} (1975) 99.
  %%CITATION = NUPHA,B88,99;%%

%\cite{Grisaru:1975bx}
\bibitem{Grisaru:1975bx}
  M.~T.~Grisaru, P.~van Nieuwenhuizen and C.~C.~Wu,
  %``Gravitational Born Amplitudes And Kinematical Constraints,''
  Phys.\ Rev.\  D {\bf 12} (1975) 397.
  %%CITATION = PHRVA,D12,397;%%




  
  %\cite{Anchordoqui:2009mm}
\bibitem{Anchordoqui:2009mm}
  L.~A.~Anchordoqui, H.~Goldberg, D.~Lust, S.~Nawata, S.~Stieberger and T.~R.~Taylor,
  %``LHC Phenomenology for String Hunters,''
  Nucl.\ Phys.\  B {\bf 821}, 181 (2009)
  [arXiv:0904.3547 [hep-ph]].
  %%CITATION = NUPHA,B821,181;%%







%\cite{Cheng:2010pt}
\bibitem{Cheng:2010pt}
  H.~C.~Cheng,
  %``2009 TASI Lecture -- Introduction to Extra Dimensions,''
  arXiv:1003.1162.
  %%CITATION = ARXIV:1003.1162;%%

%\cite{vanDam:1970vg}
\bibitem{vanDam:1970vg}
  H.~van Dam and M.~J.~G.~Veltman,
  %``Massive And Massless Yang-Mills And Gravitational Fields,''
  Nucl.\ Phys.\  B {\bf 22} (1970) 397.
  %%CITATION = NUPHA,B22,397;%%
%\cite{Zakharov:1970cc}
%\bibitem{Zakharov:1970cc}
  V.~I.~Zakharov,
  %``Linearized gravitation theory and the graviton mass,''
  JETP Lett.\  {\bf 12} (1970) 312
  [Pisma Zh.\ Eksp.\ Teor.\ Fiz.\  {\bf 12} (1970) 447].
  %%CITATION = ZFPRA,12,447;%%

%\cite{Amsler:2008zzb}
\bibitem{Amsler:2008zzb}
  C.~Amsler {\it et al.}  [Particle Data Group],
  %``Review of particle physics,''
  Phys.\ Lett.\  B {\bf 667}, 1 (2008).
  %%CITATION = PHLTA,B667,1;%%



%\cite{Hannestad:2001jv}
\bibitem{Hannestad:2001jv}
  S.~Hannestad and G.~Raffelt,
  %``New supernova limit on large extra dimensions,''
  Phys.\ Rev.\ Lett.\  {\bf 87}, 051301 (2001)
  [arXiv:hep-ph/0103201].
  %%CITATION = PRLTA,87,051301;%%

%\cite{Hannestad:2001xi}
\bibitem{Hannestad:2001xi}
  S.~Hannestad and G.~G.~Raffelt,
  %``Stringent neutron-star limits on large extra dimensions,''
  Phys.\ Rev.\ Lett.\  {\bf 88}, 071301 (2002)
  [arXiv:hep-ph/0110067].
  %%CITATION = PRLTA,88,071301;%%






%\cite{Kachelriess:2000cb}
\bibitem{Kachelriess:2000cb}
  M.~Kachelriess and M.~Plumacher,
  %``Ultrahigh energy neutrino interactions and weak-scale string theories,''
  Phys.\ Rev.\  D {\bf 62}, 103006 (2000)
  [arXiv:astro-ph/0005309];
  %%CITATION = PHRVA,D62,103006;%%
%\cite{Kachelriess:2001jq}
%\bibitem{Kachelriess:2001jq}
  M.~Kachelriess and M.~Plumacher,
  %``Remarks on the high-energy behaviour of cross-sections in weak-scale
  %string theories,''
  arXiv:hep-ph/0109184.
  %%CITATION = HEP-PH/0109184;%%

%\cite{Grzadkowski:2007zz}
\bibitem{Grzadkowski:2007zz}
  B.~Grzadkowski and J.~F.~Gunion,
  %``KK gravitons and unitarity violation in the Randall-Sundrum model,''
  Phys.\ Lett.\  B {\bf 653}, 307 (2007);
  %%CITATION = PHLTA,B653,307;%%
%\cite{Grzadkowski:2005tx}
%\bibitem{Grzadkowski:2005tx}
  %B.~Grzadkowski and J.~Gunion,
  %``Tree-level unitarity in the presence of warped geometries,''
  Acta Phys.\ Polon.\  B {\bf 36}, 3513 (2005).
  %%CITATION = APPOA,B36,3513;%%


\end{thebibliography}
\end{document}